\documentclass{ar-1col-S2O}
\usepackage[numbers]{natbib}
\usepackage{url}
\usepackage{amsmath,amssymb}

\begin{document}

% Page header
\markboth{Dalton et al.}{Memory and Friction}

% Title
\title{Memory and Friction: From the Nanoscale to the Macroscale}

\author{Benjamin A. Dalton$^1$, Anton Klimek$^1$, Henrik Kiefer$^1$, Florian N. Br\"{u}nig$^1$, H\'{e}l\`{e}ne Colinet$^1$, Lucas Tepper$^1$, Amir Abbasi$^1$, and Roland R. Netz$^1$
\affil{$^1$Freie Universität Berlin, Fachbereich Physik, 14195 Berlin, Germany; email: rnetz@physik.fu-berlin.de}
}

% ------------------------ Author information ------------------------  

% Benjamin A. Dalton
% 	email: dalton@zedat.fu-berlin.de
% 	Orcid: 0000-0002-0448-2989

% Anton Klimek
% 	email: kanton@zedat.fu-berlin.de
% 	Orcid: -

% Henrik Kiefer
% 	email: henrik.kiefer@fu-berlin.de
% 	Orcid: 0009-0008-0841-2217

% Florian N. Brünig
% 	email: florian.bruenig@fu-berlin.de
% 	Orcid: 0000-0001-8583-6488

% Helene Colinet
% 	email: helene.a.colinet@gmail.com
% 	Orcid: -

% Lucas Tepper
% 	email: l.tepper@fu-berlin.de
% 	Orcid: -

% Amir Abbasi
% 	email: amir.abbasi@fu-berlin.de
% 	Orcid: 0000-0002-0441-963X

% Roland R. Netz
% 	email: rnetz@physik.fu-berlin.de
% 	Orcid: 0000-0003-0147-0162

% -----------------------------------------------------------------------  

%Abstract
\begin{abstract}
Friction is a phenomenon that manifests across all spatial and temporal scales, from the molecular to the macroscopic scale. It describes the dissipation of energy from the motion of particles or abstract reaction coordinates and arises in the transition from a detailed molecular-level description to a simplified, coarse-grained model. It has long been understood that time-dependent (non-Markovian) friction effects are critical for describing the dynamics of many systems, but that they are notoriously difficult to evaluate for complex physical, chemical, and biological systems. In recent years, the development of advanced numerical friction extraction techniques and methods to simulate the generalized Langevin equation have enabled exploration of the role of time-dependent friction across all scales. We discuss recent applications of these friction extraction techniques and the growing understanding of the role of friction in complex equilibrium and non-equilibrium dynamic many-body systems
\end{abstract}

%Keywords, etc.
\maketitle

%Table of Contents
% \tableofcontents

% Heading 1
\section{Introduction}\label{Intro}

Many systems across a wide range of spatial and temporal scales, whether physical, chemical, biological, ecological, economic, financial, or meteorological, are many-body, interacting systems. Describing the dynamics of the individual parts is often computationally or experimentally impossible. Therefore, the focus is usually on the dynamics of a low-dimensional collective variable or reaction coordinate, typically influenced by the rest of the system not under direct observation. Examples include the motion of a tracer particle within a liquid \cite{Zwanzig_1970, Espanol_1993, Bocquet_1994, Mason_1995, Crocker_2000, Lesnicki_2016}, the vibrational modes of a molecule in gas or liquid phases \cite{Straub_1987, Berne_1990, Tuckerman_1993}, the system polarization in spectroscopy \cite{Heyden_2010, Lesnicki_2018, Carlson_2020}, the shear strain as a function of shear stress in rheology experiments \cite{Furst_2017, Schmidt_2024}, the distance between two fluorescently labeled amino acids in protein folding experiments \cite{Schuler_2008, Chung_2012, Chung_2013, Neupane_2016, Neupane_2016B}, and the position of a moving organism in biological motility studies \cite{Ridley_2003, Maiuri_2012, Mitterwallner_2020, klimek2024data}. Overall, the goal is to replace the full description of the many-body system with a low-dimensional observable and to derive accurate equations of motion that model the dynamics of that observable by coupling it with its environment's dynamics.

%\begin{marginnote}[]
%\entry{Projection}{Systematic method to single out the interesting part of phase space in the form of a one- or low-dimensional reaction coordinate, where the phase space consists of the positions
%and momenta of all atoms of the system }
%\entry{Generalized Langevin equation (GLE)}{Equation of motion for a general reaction coordinate that includes non-Markovian friction and a force due to interactions with the environment}
%\entry{Coarse graining}{Technique that reduces the number of degrees used to describe the structure and dynamics of a system}
%\entry{Non-Markovian friction}{Friction that not only depends on the current velocity of the reaction coordinate but also on its past velocities}
%\entry{Memory kernel}{The function $\Gamma(t)$ encapsulating time-dependent friction due to dissipative relaxation effects in the environment}
%\end{marginnote} 

\begin{marginnote}[]
\entry{Projection}{Systematic method to single out interesting parts of high-dimensional phase space in the form of a one- or low-dimensional observable}
\entry{Generalized Langevin equation (GLE)}{Non-Markovian equation of motion for describing the dynamics of a general stochastic observable}
\entry{Coarse graining}{Techniques that reduce the number of degrees of freedom used to describe the structure and dynamics of a system}
\entry{Non-Markovian friction}{Friction force that depends on the current and past velocities of an observable}
\entry{Memory kernel}{The function $\Gamma(t)$ encapsulating time-dependent friction due to dissipative relaxation effects in the environment}
\end{marginnote} 

The process of deriving the equation of motion for an arbitrary low-dimensional observable is known as projection \cite{Nakajima_1958, Mori_1965, Mori_1965b, Zwanzig_1961}. An appropriate observable that captures the system's essential behavior or is experimentally measurable is selected, and then a projection operator is constructed to single out the dynamics of the observable while the remaining degrees of freedom form the environment. The result is a generalized Langevin equation (GLE), which features a memory-dependent friction, a potential of mean force term, and a complementary force term (typically called a random force), representing interactions with the environment. Note that there are many choices for the projection operator \cite{Mori_1965, Zwanzig_1961, carof_two_2014, Ayaz_2022, Vroylandt_2022b, Vroylandt_2022, Vroylandt_2022c} that produce different GLEs, exact when no approximations are applied to the complementary force term. Although friction appears explicitly, the GLE is time-reversible if no additional approximations are made. The GLE parameters can be extracted from experimental or simulation time series data, and many techniques for extracting friction memory kernels are available \cite{Berne_1970, Lange_2006, Jung_2017, Daldrop_2018, Vroylandt_2022c}. Unavoidable data discretization in time can be addressed with refined estimation methods \cite{Tepper_2024}. Overall, the GLE has emerged as a common and practical coarse-graining technique \cite{chorin_optimal_2000, Darve_2009, Straube_2020, Schilling_2022}.

In this review, we cover a broad range of topics where the effects of non-Markovian friction significantly influence the system dynamics. These include the free diffusion of particles and the motility of biological cells in complex environments, the motion of reaction coordinates in confining potentials and the dynamics of rare events, the folding of proteins, molecular vibrational spectroscopy, the dynamics of meteorological weather patterns, and more. We analyze these systems from the perspective of the GLE and discuss applying memory kernel extraction to evaluate friction and dissipation time scales. Additionally, we discuss Markovian embedding simulation techniques for efficient GLE simulation, parameterized using extraction techniques. These Markovian embedding techniques consist of transforming the GLE into a set of coupled ordinary Langevin equations that can be efficiently simulated to reproduce GLE dynamics. Perturbing model parameters near extracted values efficiently probes the friction and potential energy landscape's influence on observable dynamics \cite{Ayaz_2021, Brunig_2022d}. In its original formulation, the GLE is not suitable for describing systems that are out of equilibrium in their stationary states. Living systems, such as cells and organisms, are examples of such truly non-equilibrium systems and have been intensely studied in recent years \cite{Lau_2003, Gnesotto_2018, Abbasi_2023}. The formulation of GLEs for out-of-equilibrium systems is an active field of research \cite{Schilling_2022, Schilling_2021, Netz_2024}, as we discuss briefly throughout the review.

This review is organized as follows. In Section \ref{Section_Examples}, we provide an overview of systems that have recently been analyzed using the GLE and which exhibit significant non-Markovian friction effects. In Section \ref{Section_GLE}, we provide a brief derivation of the GLE, memory-kernel extraction techniques, and the Markovian embedding framework for simulating GLEs. In Section \ref{friction_across_scales}, we present extracted memory-dependent friction kernels for three systems discussed in Section \ref{Section_Examples} and compare the system diffusivities. In Section \ref{features}, we describe three implications of the effects of memory-dependent friction on standard statistical mechanical observables. Finally, in Section \ref{applications}, we describe three investigations based on the application of the GLE: protein folding at the nanoscale (Section \ref{app_protein}), cell-type classification at the microscale (Section \ref{app_organism}), and weather pattern prediction at the macroscale (Section \ref{app_weather}).

\begin{marginnote}[]
\entry{Memory kernel extraction}{Methods to determine $\Gamma(t)$ from time series data}
\entry{Markovian embedding}{Method to transform the non-Markovian GLE into a set of coupled Markovian Langevin equations}
\end{marginnote} 

\begin{figure}[t!]
\includegraphics[width=6in]{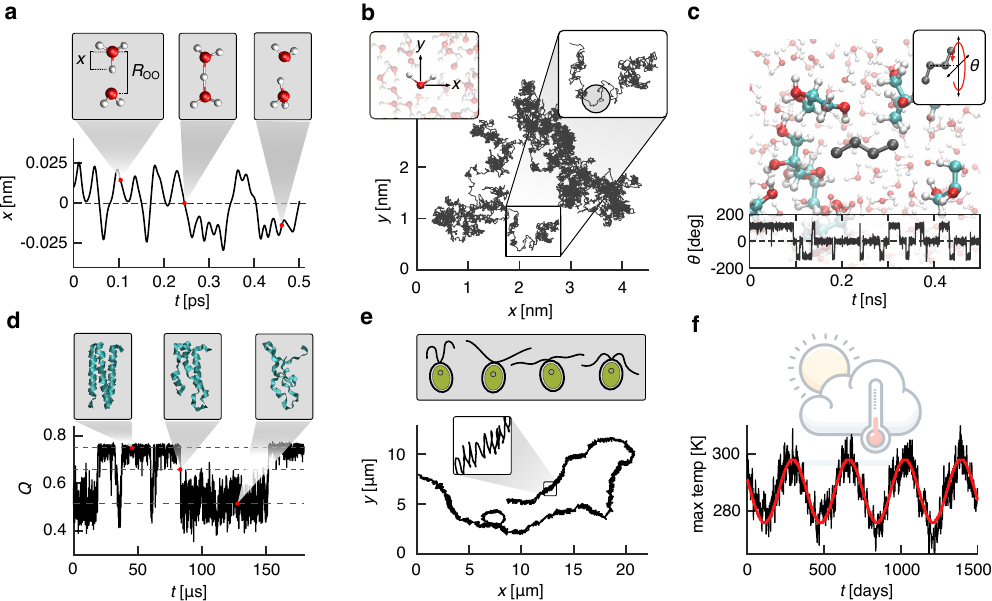}
\caption{From nano to macro: dynamics across different spatial and temporal scales. (\textit{a}) An ab initio molecular dynamics (MD) simulation of the proton-transfer reaction between two water molecules at a fixed relative distance of R$_{\mathrm{OO}}=2.56 \textup{~\AA}$. The reaction coordinate $x(t)$ represents the excess proton’s position along the oxygen-oxygen axis. \cite{Brunig_2022c}. (\textit{b}) Classical MD simulation of the diffusion of a single water molecule in a liquid water environment, showing the 1 ns path of the oxygen atom in the $x,y$ plane. The magnification focuses on a 50 ps trajectory segment where the short-time-scale inertia- and memory-dominated dynamics can be observed (gray circle). (\textit{c}) Isomerization dynamics of an n-butane molecule solvated in a mixed water-glycerol solvent (20$\%$ mass fraction glycerol), simulated using classical MD. The time evolution of the n-butane dihedral angle $\theta(t)$ is shown, as it transitions between distinct isomeric states. The composition of the complex solvent environment influences the dynamics of dihedral isomerization \cite{Dalton_2024}. (\textit{d}) The folding dynamics of the 73-residue $\alpha_3$D fast-folding protein. The fraction of native contacts reaction coordinate $Q(t)$ is a one-dimensional collective-coordinate representation of the full atomic structure of the protein, which transitions back and forth between folded and unfolded states. Representative molecular structures of a folded state and unfolded state, as well as a transition state, are shown \cite{Dalton_2023}. (\textit{e}) An 8-second swimming trajectory of the center of a \textit{Chlamydomonas reinhardtii} algae cell. The illustrated stroke cycle takes approximately 25 ms, and the persistent oscillations from the stroking motion are visible in the magnification. On longer timescales, the cell executes a random walk \cite{klimek2024data}. (\textit{f}) The daily maximal temperature in Berlin-Tegel is plotted over four years. The temperature exhibits large fluctuations on the daily time scale as well seasonal oscillations. Predicting daily temperature is of practical interest.}
\label{Figure_1}
\end{figure}

\section{Examples of systems that are governed by non-Markovian friction}\label{Section_Examples}

In Figure~\ref{Figure_1}, we show an overview of different systems discussed throughout this review. These systems span the range from the nanoscopic scale of a quantum-chemical description of subatomic particle transport to the scale of biological cells, and up to the terrestrial scale of meteorological processes. In all of these examples, the application of GLE modeling and friction memory-kernel extraction has illuminated the underlying dynamics of the relevant observables, thereby enhancing our understanding of the holistic dynamics of these systems. Figure~\ref{Figure_1}\textit{a} illustrates the proton-transfer reaction between two water molecules at a fixed separation, obtained from ab initio molecular dynamics (MD) simulations. The dynamics of excess protons in water is important in chemistry and biology and plays a fundamental role in acid-base reactions, enzyme catalysis, and proton transport in biological membranes \cite{Wraight2006, Agmon2016}. This process exhibits a significantly higher diffusion coefficient than other ions in water. This distinct property is due to the Grotthuss mechanism, by which excess protons subsequently transfer between individual water molecules and exchange identities with other hydrogen nuclei within hydronium ions \cite{Agmon2016, Agmon1995, Marx2006}. An elementary step in the long-distance diffusion of protons in water is therefore the barrier crossing between two water molecules \cite{Thamer2015, Dahms2017, Roy2020, Brunig_2022, Brunig_2022c}. In Figure~\ref{Figure_1}\textit{b}, we see the diffusive transport of a single water molecule in bulk liquid water. The trajectory reveals self-similarity, a characteristic of diffusion that breaks down below the picosecond scale, where inertia and memory effects dominate the molecular dynamics \cite{Kowalik_2019, Straube_2020, Scalfi_2023}. In Figure~\ref{Figure_1}\textit{c} and \ref{Figure_1}\textit{d}, we highlight the importance of molecular conformation dynamics. In Figure~\ref{Figure_1}\textit{c}, the isomerization dynamics of a single dihedral bond in the n-butane molecule is strongly influenced by the complex mixed water-glycerol solvent environment. Simulations of n-butane, and other small, isomerizing molecules, are computationally inexpensive and provide an ideal system for investigating internal friction effects and the role of viscogenic coupling in small isomerizing molecules \cite{Daldrop_2018, Yamaguchi_2021, Dalton_2024}. The folding and unfolding dynamics of the fast-folding $\alpha_3$D protein shown in Figure~\ref{Figure_1}\textit{d}, however, are significantly computationally expensive and such classical MD simulations require advanced special-purpose computers \cite{Lindorff_2011, Shaw_2009}. The study of the friction acting on a folding protein has a long history, particularly due to the pronounced influence of internal friction on folding dynamics \cite{Beece_1980, Doster_1983, Ansari_1992, Jas_2001, Hagen_2010, Soranno_2012, Borgia_2012}. Extensive trajectories, like those in Figure~\ref{Figure_1}\textit{d}, provide a crucial testing ground to study memory friction effects in protein folding \cite{Dalton_2023}. In Figures~\ref{Figure_1}\textit{e} and \ref{Figure_1}\textit{f}, we emphasize systems at greater spatial and temporal scales. The swimming \textit{Chlamydomonas reinhardtii} alga in Figure~\ref{Figure_1}\textit{e} must overcome strong hydrodynamic resistance to move through its fluid-medium environment. At short time scales, its motion is characterized by the periodic beating of its flagella. At longer time scales, it appears to execute undirected diffusive motion. The motion of organisms across multiple time scales has long been described using stochastic models \cite{Klafter_2005, bartumeus2005animal, johnson2008continuous, dieterich2008anomalous, viswanathan2011physics}, such as Lévy random walks \cite{Viswanathan_1996, edwards2007revisiting} and even GLEs \cite{Mitterwallner_2020, klimek2022optimal,klimek2024data}. In this review, we will discuss a novel application of the GLE as a classification scheme for determining modes of cellular motion. In Figure~\ref{Figure_1}\textit{f}, and in the final section of this review, we explore the terrestrial scale. Coarse graining is absolutely necessary for describing terrestrial phenomena, such as the atmosphere, economies, financial systems, or ecosystems, since it is impossible to explicitly model the myriad interacting agents in detail. Weather patterns, such as the daily maximal temperature measured at Berlin-Tegel, as presented in Figure~\ref{Figure_1}\textit{f}, follow long-term seasonal trends and short-term stochasticity. By filtering out the deterministic modes in the signal, the remaining stochastic process can be modeled by the GLE, which provides a framework for highly accurate daily temperature predictions with a fraction of the computational cost of machine-learning techniques.

\section{Memory-dependent friction and the generalized Langevin equation (GLE)}\label{Section_GLE}

This review discusses recent applications of the GLE as a coarse-graining tool for extracting and analyzing time-dependent friction memory kernels from time series data across various physical, chemical, and biological systems. In this section, we provide brief derivations of the GLE, the friction memory-kernel extraction method, and the Markovian embedding approach for simulating the GLE.
 
\subsection{Derivation of the GLE: a sketch}\label{GLE_Derivation}

The derivation of the GLE is based on the dynamics of a general many-body system that is fully characterized by the Hamiltonian $ {\cal H}(R,P)$ in terms of the $3N$ particle, or atomic, positions $R$ and their $3N$ momenta $P$. The Liouville operator ${\cal L}(R,P)$, which depends on the Hamiltonian, gives the rate of change of an arbitrary phase-space-dependent observable $x(R,P,t)$ as $\dot{x}(R,P,t) = {\cal L}(R,P) x(R,P,t)$. The solution of this differential equation can be formally written using the operator exponential as $x(R,P,t) = \exp( t {\cal L}(R,P)) x(R,P,0)$, from which the acceleration of the observable is obtained by differentiation as $ \ddot x(R,P,t) = \exp( t {\cal L}(R,P)) {\cal L}^2(R,P)x(R,P,0)$. A crucial step in the derivation of the GLE is introducing a projection operator ${\cal P}$ to single out a relevant part of phase space, such as the observable $x(R,P,0)$ itself, its rate of change $\dot x(R,P,0)$, or a combination of both. Together with the complementary projection operator ${\cal Q}$, the identity operator $1 = {\cal P} + {\cal Q}$ is obtained, which defines ${\cal Q}$. The starting point of the derivation of the GLE is to insert this identity operator into the expression for the acceleration of the reaction coordinate according to
\begin{equation}\label{GLE0}
\begin{split}
\ddot{x}(R,P, t) &= e^{ t {\cal L}(R,P)} {\cal L}^2(R,P) ({\cal P}+ {\cal Q}) x(R,P,0) \\
 &= e^{ t {\cal L}(R,P)}  {\cal L}^2(R,P) {\cal P}  x(R,P,0)  +  e^{ t {\cal L}(R,P)}   {\cal L}^2(R,P)  {\cal Q}  x(R,P,0),
 \end{split}
\end{equation}
where in the last step, we used the fact that both the exponential and Liouville operators are linear. The acceleration $\ddot{x}(R,P, t)$ splits into two parts. The first is produced by the relevant projection operator ${\cal P}$ and gives rise to forces from the gradient of a conservative potential, as described by Newton's equation of motion. The second part comes from the complementary projection operator ${\cal Q}$ and accounts for forces due to environmental interactions. After a few additional steps of manipulation, one arrives at the GLE in the general form \cite{Ayaz_2022, Vroylandt_2022b,Netz_2024}
\begin{equation}\label{GLE_General}
\ddot{x}(t) = -\nabla \bar{U}\big[x(t)\big] -\int\limits_{0}^{t}\bar{\Gamma}(t-t^{\prime})\dot{x}(t^{\prime})dt^{\prime} + \bar{F}_{\text{R}}(t),
\end{equation}
which consists of a potential gradient term $\nabla \bar{U}(x)$, a friction term with friction memory kernel $\bar{\Gamma}(t)$, and a complementary force $\bar{F}_{\text{R}}(t)$, which is often interpreted as a stochastic force. 
The force correlator approximately follows $\langle \bar{F}_{\text{R}}(t) \bar{F}_{\text{R}}(t') \rangle = \langle \dot x^2 \rangle\bar{\Gamma}(t-t')$ 
\cite{Ayaz_2022, Vroylandt_2022b, Vroylandt_2022,AyazTurkish2022}. 
 There are many different projection operators ${\cal P}$, which lead to different GLEs. For details, see \cite{Mori_1965, Mori_1965b, Zwanzig_1961, Ayaz_2022, Vroylandt_2022b, Vroylandt_2022, AyazTurkish2022}. Note that, in writing Eq.~\ref{GLE_General}, we dropped the phase-space dependence of the observable $x(t)$ and the complementary force $\bar{F}_{\text{R}}(t)$. The overbar notation indicates a GLE formulation where no mass appears in front of the acceleration, which applies to both equilibrium and non-equilibrium systems. Consequently, the quantities $\bar{U}(x)$, $\bar{\Gamma}(t)$ and $\bar{F}_{\text{R}}(t)$ do not have their usual physical units.

For equilibrium systems, where a heat-bath temperature is defined, the equipartition theorem $m\langle \dot x^2 \rangle = k_{\text{B}}T$ holds, allowing the introduction of an effective mass $m$, which is specific to the observable. Thus, Eq.~\ref{GLE_General} can be written in a standard equilibrium form
\cite{Daldrop_2018, Kowalik_2019, Ayaz_2021, Brunig_2022a, Brunig_2022d, Dalton_2023, Dalton_2024}
\begin{equation}\label{GLE}
m\ddot{x}(t) = -\nabla U\big[x(t)\big] -\int\limits_{0}^{t}\Gamma(t-t^{\prime})\dot{x}(t^{\prime})dt^{\prime} + F_{\text{R}}(t),
\end{equation}
where the quantities ${U}(x)$, ${\Gamma}(t)$, and ${F}_{\text{R}}(t)$ have their usual physical units. Note that the mass $m$ and the friction kernel ${\Gamma}(t)$ can, in principle, depend on $x$ \cite{Ayaz_2022, AyazTurkish2022}, which is, however, not covered in this review. For simplicity, we only consider one-dimensional observables; generalizing to vectorial observables is possible and relevant but not explicitly discussed here. When discussing equilibrium systems throughout this review, we use Equation~\ref{GLE}. For non-equilibrium systems, such as cellular motility (Sections \ref{friction_across_scales} and \ref{app_organism}) and meteorological processes (Section \ref{app_weather}), we maintain the general form of Equation~\ref{GLE_General}.

\subsection{Extraction of the GLE parameters from time-series data}\label{GLE_Extraction}

Recent advances in the study of memory-dependent friction effects in complex systems have become possible through the development of robust memory kernel extraction techniques \cite{Berne_1970, Daldrop_2018, Kowalik_2019, Ayaz_2021}. With sufficient time series data, the observable's stochastic trajectory can be mapped onto a GLE, and a numerical representation for the friction memory kernel $\Gamma(t)$ is evaluated. In this section, we derive an extraction scheme for the GLE as given in Eq.~\ref{GLE}.

Recently, techniques have focused on extracting the running integral of the memory kernel $G(t)$, defined as
\begin{equation}
G(t) = \int_0^{t}\Gamma(t')dt' ,
\end{equation}
such that a distinct plateau identifies the friction coefficient, $\gamma$, for the observable, defined as $\gamma \equiv G(t \rightarrow \infty)$ \cite{Kowalik_2019, Ayaz_2021}. Equivalent methods can be used for directly extracting $\Gamma(t)$ \cite{Kowalik_2019, Daldrop_2018}. To extract a numerical approximation to $G(t)$, we multiply the generalized Langevin equation (GLE) in Eq.~\ref{GLE} by the initial velocity $\dot{x}(0)$ and evaluate the ensemble average, leading to the correlation function $C^{\dot{x}\ddot{x}}(t) = \big\langle \dot{x}(0)\ddot{x}(t) \big\rangle$. Since the random force in Eq.~\ref{GLE} lies in a subspace orthogonal to $\dot{x}(t)$ for all $t$, both $\big\langle \dot{x}(0)F_{\text{R}}(t)\big\rangle =0$ and $\big\langle {x}(0)F_{\text{R}}(t)\big\rangle =0$. Integrating the entire equation once, we arrive at the relationship
\begin{equation}\label{G_Relation}
\frac{C^{\dot{x}\dot{x}}(t)}{C^{\dot{x}\dot{x}}(0)} C^{{x}\nabla U}(0)= C^{{x}\nabla U}(t) -\int_{0}^{t}G(t-t^{\prime})C^{\dot{x}\dot{x}}(t^{\prime})dt^{\prime},
\end{equation}
see \cite{Ayaz_2021} for details. 

Eq.~\ref{G_Relation} is expressed in terms of two correlation functions, $C^{\dot{x}\dot{x}}(t) = \big\langle \dot{x}(0)\dot{x}(t) \big\rangle $ and $C^{{x}\nabla U}(t) = \big\langle {x}(0)\nabla U\big[x(t)\big] \big\rangle $, both of which are easily evaluated from long time series trajectories. A key feature of this memory kernel extraction technique is that it applies generally to systems evolving across arbitrary free-energy landscapes $U(x)$. For the time series trajectories shown in Figures~\ref{Figure_1}\textit{a}, \ref{Figure_1}\textit{c}, and \ref{Figure_1}\textit{d}, the free energy profiles $U(x)$ are extracted using $U(x) = -k_{\text{B}} T\log\big[\rho(x)\big]$, where $\rho(x)$ is the probability density over $x(t)$, evaluated via histogram binning, and $\nabla U\big[x(t)\big]$ is constructed by interpolation. Eq.~\ref{G_Relation} can be discretized using the trapezoidal integration rule, leading to 
\begin{equation}\label{G_Relation_disc}
\frac{C^{\dot{x}\dot{x}}_i}{C^{\dot{x}\dot{x}}_0} C^{{x}\nabla U}_0= C^{{x}\nabla U}_i - \frac{\Delta t }{2}G_iC^{\dot{x}\dot{x}}_0 - \Delta t \sum_{j=1}^{i-1}G_{i-j}C^{\dot{x}\dot{x}}_j.
\end{equation}
Using $G(0)=G_0=0$, we solve for $G_i = G(t)$ and arrive at the iterative scheme
\begin{equation}\label{It_G}
G_i = \left\{ 
 \begin{array}{l l l l l}
 0, & \quad i=0\\
 \quad \\
     \frac{2}{\Delta tC^{\dot{x}\dot{x}}_{0}}\bigg[C_1^{\nabla Ux} - \frac{C_0^{\nabla Ux}}{C_0^{\dot{x}\dot{x}}}C_1^{\dot{x}\dot{x}} \bigg], & \quad i=1 \\
  \quad \\      
        \frac{2}{\Delta tC^{\dot{x}\dot{x}}_{0}}\bigg[C_i^{\nabla Ux} - \frac{C_0^{\nabla Ux}}{C_0^{\dot{x}\dot{x}}}C_i^{\dot{x}\dot{x}} - \Delta t\sum_{j=1}^{i-1}G_{j}C_{i-j}^{\dot{x}\dot{x}} \bigg]. & \quad i>1
  \end{array} \right.
\end{equation}
The discretized form of the memory kernel, $\Gamma_i$, can be obtained through numerical differentiation.

\subsection{Simulation of the GLE via Markovian embedding}\label{GLE_Simulation}
	
Complementary to memory kernel extraction techniques discussed in the previous section, Markovian embedding techniques offer an efficient way to simulate a GLE even for long-ranged memory kernels without the need to perform the numerically demanding integral over $\Gamma(t)$ in Eq.~\ref{GLE}. This is achieved by introducing additional degrees of freedom, which map the GLE in Eq.~\ref{GLE} onto a system of linearly coupled standard Langevin equations \cite{Risken_1996, Ceriotti_2010}. To see this, consider the system of coupled equations \cite{Zwanzig_1973, Kappler_2019, Ayaz_2021} 
\begin{equation}\label{eq_LE}
\begin{split}
    m\ddot{x}(t) &= -\frac{\mathrm{d} U(x)}{\mathrm{d} x} - \sum_{n=1}^N \frac{\gamma_n}{\tau_n} \left(x(t)-y_n(t)\right), \\
    \dot{y}_n(t) &= -\frac{1}{\tau_n}\left( y_n(t) - x(t) \right) + \sqrt{\frac{k_{\text{B}} T}{\gamma_n}}\eta_n(t)  \quad \text{ for } n=1,2,\dots, N,
\end{split}
\end{equation}
where $U(x)$ is an external potential that only affects the observable $x$. Each auxiliary degree of freedom $y_n(t)$ is subject to stationary Gaussian noise $\eta_n(t)$ with $\langle \eta_n(t) \rangle = 0$ and $\langle \eta_n(t), \eta_l(t^\prime)\rangle = 2\delta_{nl}\delta(t-t^\prime)$. Eq.~\ref{eq_LE} can be mapped onto a GLE for the observable $x(t)$. This involves solving the inhomogeneous first-order differential equation for the variables $y_n(t)$ and inserting the solution into the Langevin equation for $x(t)$. The solution for $y_n(t)$ reads
\begin{equation}\label{eq_yn}
   y_n(t) = \left( y_n(0) - x(0) \right) e^{-t/\tau_n} + x(t) - \int_0^t\mathrm{d}s\,e^{-(t-s)/\tau_n} \dot{x}(s) + \int_0^t\mathrm{d}s\,e^{-(t-s)/\tau_n}\frac{1}{\gamma_n}\eta_n(s).
\end{equation}
Inserting Eq.~\ref{eq_yn} into Eq.~\ref{eq_LE} returns the original GLE Eq.~\ref{GLE} with $F_{\text{R}}(t) = \sum_n\left[ \frac{\gamma_n}{\tau_n}(q(0)-y_n(0))e^{-t/\tau_n}- \frac{1}{\tau_n}\int_0^t\mathrm{d}s,e^{(t-s)/\tau_n}\eta_n(s)\right]$ and the multi-exponential kernel $\Gamma(t) = \sum_n\frac{\gamma_n}{\tau_n}e^{-t/\tau_n}$. Assuming that the system in Eq.~\ref{eq_LE} is initially in equilibrium, the initial values $y_n(0)$ and $x(0)$ are distributed according to a Boltzmann distribution \cite{Risken_1996}. Using this, it follows that $\langle F(t) \rangle=0$ and $\langle F(t) F(0)\rangle = k_{\text{B}}T\Gamma(t)$. Hence, we obtain a GLE with a memory kernel consisting of a sum of exponentially decaying functions. One can generate a trajectory of the GLE in Eq.~\ref{GLE} by numerically integrating the coupled system of ordinary Langevin equations in Eq.~\ref{eq_LE}. It is also important to note that memory kernels with non-exponential basis functions, such as decaying-oscillating functions, can be represented by slight modifications of Eq.~\ref{eq_LE} \cite{Brunig_2022a, Brunig_2022d, Dalton_2024}.

\section{Memory kernels and diffusivity}\label{friction_across_scales}

\begin{figure}[t!]
\includegraphics[width=6 in]{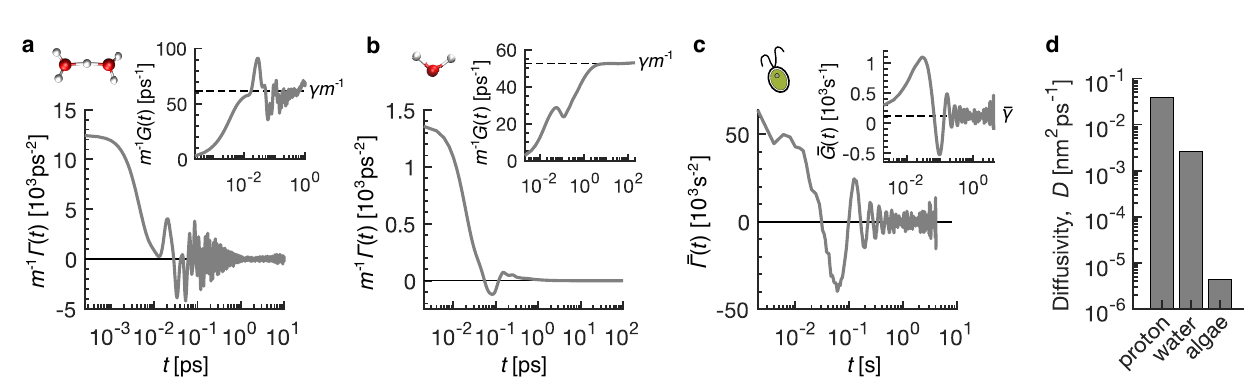}
\caption{Friction at various spatial and temporal scales. (\textit{a}) The memory kernel $\Gamma(t)$ and its running integral $G(t) = \int_0^{t}\Gamma(t^{\prime})dt^{\prime}$ for the proton-transfer dynamics presented in Figure~\ref{Figure_1}\textit{a}. The plateau value $G(t\rightarrow\infty) = \gamma$ evaluates the friction coefficient $\gamma$. $\Gamma(t)$, $G(t)$, and $\gamma$ (black dashed line) are scaled by $m^{-1}$ to facilitate comparison with non-equilibrium systems, which can be described using Eq.~\ref{GLE_General}. (\textit{b}) The friction on the center of mass of an individual water molecule in bulk liquid water (see Figure~\ref{Figure_1}\textit{b}). (\textit{c}) The memory kernel $\bar{\Gamma}(t)$ and running integral $\bar{G}(t) = \int_0^{t}\bar{\Gamma}(t^{\prime})dt^{\prime}$ for a living and swimming \textit{Chlamydomonas reinhardtii} algae cell (see Figure~\ref{Figure_1}\textit{e}). The plateau value $\bar{G}(t\rightarrow\infty) = \bar{\gamma}$ is used to evaluate the friction coefficient $\bar{\gamma}$. (\textit{d}) The diffusivity $D=\langle \dot x^2\rangle \tau_{\text{m}}$ compared for the three systems, where $ \tau_{\text{m}} = m\gamma^{-1}$ for the proton and the water molecule, and $ \tau_{\text{m}} = \bar{\gamma}^{-1}$ for the algae cell. For equilibrium systems, where the equipartition theorem $\langle \dot x^2 \rangle = k_{\text{B}}T/m$ holds (see Section \ref{GLE_Derivation}), $D =k_{\text{B}}T/\gamma$.}
\label{Figure_2}
\end{figure}

In recent years, the memory extraction techniques described in Section \ref{GLE_Extraction} have been applied to a wide range of physical, chemical, and biological systems, such as those presented in Figure~\ref{Figure_1}. In each case, a one-dimensional reaction coordinate is identified, and the time series of that reaction coordinate is mapped onto a suitably chosen GLE. In many instances, the GLEs presented in either Eq.~\ref{GLE_General} or Eq.~\ref{GLE} are sufficient to model the dynamics of the reaction coordinate. In Figures~\ref{Figure_2}\textit{a}, \ref{Figure_2}\textit{b}, and \ref{Figure_2}\textit{c}, we show the memory kernels extracted for three systems presented in Figure~\ref{Figure_1}. Due to the convenience of their Cartesian-position reaction coordinates, we can use the extracted friction coefficients $\gamma$ to directly compare the diffusivities of these three systems. 

In Figure~\ref{Figure_2}\textit{a}, we show $m^{-1}\Gamma(t)$ and $m^{-1}G(t)$ for the excess proton transport presented in Figure~\ref{Figure_1}\textit{a}. This system is characterized by length scales on the order of the OH water bond length, $1\;\textup{\AA}$, and time scales on the order of the vibrational period of the water OH bond, $10\;\text{fs}$. However, the memory friction along the reaction coordinate introduces additional time scales on the order of the structural relaxation time of the water environment, which is approximately $1\;\text{ps}$ \cite{Brunig_2022a}. The kernels exhibit both oscillating and decaying behavior, and we identify a plateau value of approximately $m^{-1}G(\infty)= m^{-1}\gamma = 62\;\text{ps}^{-1}$, which can be used to evaluate the inertia time, $\tau_{\text{m}}=m/\gamma$, the timescale above which friction becomes relevant: $\tau_{\text{m}}=16\;\text{fs}$. The memory friction acting on the water molecule also exhibits a long-timescale decay (Figure~\ref{Figure_1}\textit{b}), on the order of $1$-$10\;\text{ps}$, with fewer oscillations than for the proton in Figure~\ref{Figure_2}\textit{a}, and is comparable to the relaxation timescale of water hydrogen bonding, which is approximately $8\;\text{ps}$. For the water molecule, we find $\tau_{\text{m}}=19\;\text{fs}$, similar to the value for the proton transfer reaction. However, the diffusivity, determined by the friction coefficient $\gamma$, is significantly different between these two systems due to their unequal masses. In Figure~\ref{Figure_1}\textit{c}, we show the friction acting on the \textit{Chlamydomonas reinhardtii} algal cell, $\bar{\Gamma}(t)$ and $\bar{G}(t)$, where the cell dynamics are described using Eq.~\ref{GLE_General}. The large and slowly decaying oscillations are associated with short-timescale flagella beating. For this cell, the extracted friction coefficient is $\bar{G}(\infty) = \bar{\gamma} = 0.1{\times}10^3\;\text{s}^{-1}$. Therefore, we obtain $\tau_{\text{m}}=1/\bar{\gamma} = 0.01\;\text{s}$. From the friction coefficient for each system, we compare the diffusivity $D = \langle \dot x^2\rangle \tau_{\text{m}}$ (Figure~\ref{Figure_2}\textit{d}), a suitable quantity for characterizing both equilibrium and non-equilibrium dynamics. In equilibrium, where temperature is defined and the equipartition theorem $\langle \dot x^2 \rangle = k_{\text{B}}T/m$ holds, the diffusivity follows the familiar Einstein expression $D = k_{\text{B}}T/\gamma$. The diffusivity varies considerably between the three systems, spanning four orders of magnitude. As expected, cells, due to their large size, diffuse much more slowly than protons and water molecules.

\section{Characteristic features of systems with non-Markovian friction}\label{features}

The study of molecular friction reveals characteristic signatures of non-Markovian friction effects on fundamental dynamic properties. In this section, we discuss key examples using simple model simulations, highlighting the generality of the implications for more complex systems.

\subsection{Mean squared displacement}

\begin{figure}[t]
\includegraphics[width=5in]{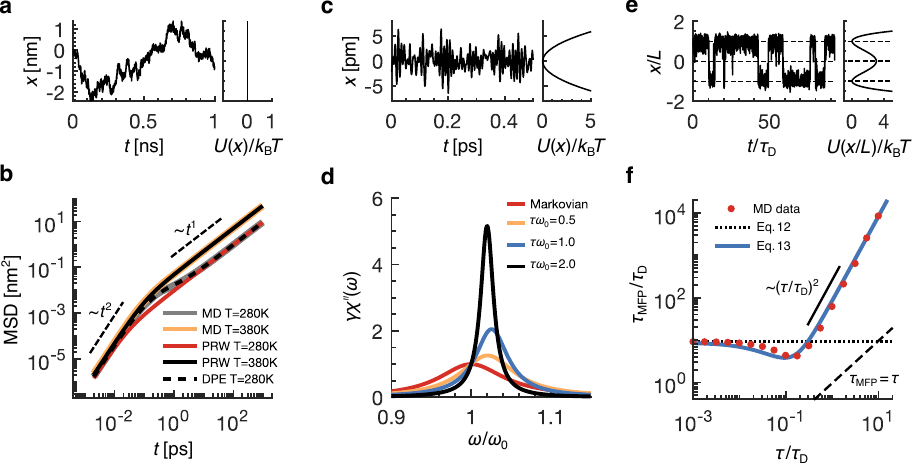}
\caption{Memory effects on fundamental dynamic system properties. (\textit{a})-(\textit{b}): The mean squared displacement (MSD) for a water molecule in bulk liquid water. (\textit{a}) The time series for the oxygen atom $x$-component for a single water molecule from a molecular dynamics (MD) simulation of bulk water at $T=280\;\text{K}$. The molecule freely diffuses in the absence of a mean force potential. (\textit{b}) MSDs are calculated from MD simulations at $T=280\;\text{K}$ (gray line) and $T=380\;\text{K}$ (yellow line). The persistent random walk model (PRW) describes the high-temperature MSD, capturing the ballistic ($\sim t^2$) and diffusive ($\sim t^1$) regimes (black line). However, it fails to capture the intermediate regime at low temperature (red line). This intermediate regime is described by the GLE with a memory kernel with a delta peak and a single exponential (delta-plus-exponential, DPE). (\textit{c})-(\textit{d}): Velocity autocorrelation function spectral analysis for non-Markovian simulations. (\textit{c}) The trajectory of a one-dimensional GLE simulation in a harmonic potential with strength $k=0.25 k_{\text{B}} T/\text{pm}^2$ for a particle with mass $m=2\;\text{u}$ and a single-exponential memory kernel $\Gamma(t) = \gamma  e^{-t/\tau}/\tau$ with $\tau \omega_0=10$. (\textit{d}) Absorption spectra according to Eq.~\ref{linres_autocorr} from GLE simulations across various memory time scales, with $\tau_\text{m}\omega_0 = 10$ constant. The spectrum for a Markovian simulation $\Tilde{\Gamma}(\omega) = \gamma$ is shown (solid red line). $\gamma \chi^{\prime\prime}(\omega_0) = 1$ for a single-exponential memory kernel, which follows from inserting the Fourier-transformed kernel $\tilde{\Gamma}(\omega)$ into Eq.~\ref{eq:linres_ft}. The systems with memory friction exhibit a blue shift and line narrowing. (\textit{e})-(\textit{f}): Non-Markovian barrier-crossing kinetics. (\textit{e}) The trajectory of a one-dimensional GLE simulation in a double-well potential $U(x) = U_0\big[(x/L)^2-1\big]^2$, with $U_0=3k_{\text{B}}T$ and characteristic length $L$. (\textit{f}) Mean first-passage times (MFPT) $\tau_{\text{MFP}}$, measuring barrier-crossing times, are shown from GLE simulations (red points). Simulations are performed with a single-exponential memory kernel over a range of memory times $\tau$. The Markovian barrier crossing time prediction Eq.~\ref{tau_Exact} is shown with the dotted black line and the non-Markovian prediction Eq.~\ref{tau_Heur} is shown with the blue line. The black dashed line for $\tau_{\text{MFP}}=\tau_{\Gamma}$ indicates that reaction kinetics are influenced by non-Markovian effects even when $ \tau \ll \tau_{\text{MFP}}$.}
\label{Figure_3}
\end{figure}

% (a), (b), and (c) The time series for an observable 

The mean squared displacement (MSD) is a fundamental measure in the study of diffusive molecular transport processes and statistical mechanics. As a time-correlation function, it measures the average distance a particle moves from its initial position over time, providing key insights into diffusion and molecular dynamics. The MSD can distinguish between different types of motion, such as Brownian or anomalous diffusion, and reveals how environmental factors like temperature or viscosity affect molecular dynamics.

\begin{marginnote}[]
\entry{Persistent random walk (PRW) model}{A stochastic model for describing the motion of particles or organisms and includes the crossover from ballistic motion at short times to diffusive motion at long times.}
\end{marginnote} 
In Figure~\ref{Figure_3}\textit{a}, we show a typical trajectory for the $x$-component of a water molecule's oxygen atom, taken from bulk liquid water simulations using classical force-field MD. In Figure~\ref{Figure_3}\textit{b}, we show the MSD for the motion of a single water molecule, simulated at two different temperatures. In a purely Markovian system with a delta-like memory kernel $\Gamma(t) = 2 \gamma \delta(t)$, one finds short-timescale ballistic dynamics with $\sim t^2$ scaling and long-timescale diffusive dynamics with $\sim t^1$ scaling. The MSD for a memoryless persistent random walk (PRW) is given by $\text{MSD}(t) =\langle (x(t)-x(0))^2\rangle = 2D[ t - \tau_m (1 - e^{-t/\tau_m})]$, where $D$ is the diffusivity introduced in the previous section. The PRW accurately describes the MSD for water molecules at $T=380\;\text{K}$ but fails to describe the water MSD at $T=280\;\text{K}$. Recently, theoretical predictions were reported for MSDs that explicitly account for memory contributions suitable for systems with simple memory kernels that satisfy the general form $\bar{\Gamma}(t) = 2a\delta(t) + b e^{-t/\tau}\cos (\Omega t)$ \cite{klimek2024data}. In Figure~\ref{Figure_3}\textit{b}, we show that these non-Markovian predictions accurately describe the data for $T=280\;\text{K}$, using $\Gamma(t)$ parameters selected to optimize the agreement between the predicted MSD and the data ($\Omega=0$, $a=27.5\;\text{ps}$, $b=46.8\;\text{ps}^{-2}$, and $\tau=1.27\;\text{ps}$). The decrease of memory effects for higher temperatures highlights that memory contributions originate from water-water interactions, such as hydrogen bonds, which exert less influence as thermal fluctuations increase. Additionally, the effects of power-law memory kernel contributions \cite{kou2004generalized, Goychuk_2009} or multi-exponential memory kernel contributions \cite{goychuk2012viscoelastic} on the MSD can be considered.

\subsection{Velocity auto-correlation functions and vibrational spectra}

\begin{marginnote}[]
\entry{Fluctuation-dissipation relation}{
Exact relation between the two-point correlation function of an observable and the linear response of that observable to an external time-dependent force.}
\end{marginnote} 

Vibrational spectroscopy is a powerful experimental tool for studying intra- and intermolecular interactions in liquids and solids. In linear absorption spectroscopy, the effective vibrational bond potential and frictional coupling between a vibrational mode and the environment are reflected in the line-shape of the absorption spectrum \cite{Brunig_2022a, Schrader_1995, Bakker2010}. The linear absorption rate of a vibrational mode is proportional to the imaginary part of the frequency-dependent susceptibility $\Tilde{\chi}''(\omega)$, which relates to the Fourier-transformed velocity autocorrelation function $C^{\dot{x}\dot{x}}(t)$ via the fluctuation-dissipation relation \cite{Kubo_1966}
\begin{equation}\label{linres_autocorr}
\omega \Tilde{\chi}''(\omega) = \frac{ \Tilde{C}^{\dot{x}\dot{x}} (\omega)}{2 k_B T}\,.
\end{equation} 
Vibrational spectroscopy covers a broad range of time scales, from collective modes and rotational motion in the GHz regime to rapid intramolecular vibrations in the high THz (i.e., the infrared) regime \cite{Carlson_2020}. Conventionally, the line shape is interpreted by a combination of homogeneous and inhomogeneous line broadening effects, which characterize adiabatic energy dissipation and nonadiabatic coupling to the environment, respectively \cite{Schrader_1995, Auer2008, Bakker2010, Perakis2016}. In contrast, in the framework of the GLE, the coupling of a vibrational mode to the environment is described by the frequency-dependent friction function, which is the Fourier-transformed time-dependent memory friction kernel, which provides a timescale bridging approach to the theory of vibrational spectroscopy. Several studies have employed the GLE to model linear absorption and vibrational relaxation by modeling vibrational modes as anharmonic oscillators coupled to a thermal bath \cite{Tuckerman_1993, Brunig_2022a, Metiu1977, Oxtoby_1981, Adelman1988, Smith1990a, Bader1996, Gottwald_2016}. Earlier works established the line-shifting and broadening principles of the GLE model but were limited by simple heuristic approximations for the friction function, which is obtained accurately using recent data-driven approaches \cite{Gottwald_2016, Brunig_2022a}.

For a harmonic potential $U(x) = \frac{k}{2} x^2$, the linear response function $\Tilde{\chi}(\omega)$, defined by the relation $\Tilde{x}(\omega)= \Tilde{\chi}(\omega) \Tilde{F}_R(\omega)$ between force and position, is obtained by Fourier transformation of the GLE in Eq.~\ref{GLE} according to $\Tilde{x}(\omega)= \int_{-\infty}^\infty dt e^{\imath \omega t} x(t)$ as \cite{Brunig_2022a, Metiu1977, Oxtoby_1981} 
\begin{equation}\label{eq:linres_ft}
\Tilde{\chi}(\omega) = \frac{1}{k-m\,\omega^2-i\omega\,\Tilde{\Gamma}(\omega)}\,.
\end{equation}
For Markovian friction $\Tilde{\Gamma}(\omega) = \gamma$, the spectral peak of $\omega\Tilde{\chi}''(\omega)$ is located at the eigenfrequency $\omega_0 = \sqrt{k/m}$, with a peak value given by $\omega_0 \Tilde{\chi}''(\omega_0) = 1/\gamma$ and a full width at half maximum (FWHM) $\sigma = 1/\tau_{\text{m}}$, where $\tau_{\text{m}} = m/\gamma$ is the inertia relaxation time, as discussed earlier. The time series trajectory in Figure~\ref{Figure_3}\textit{c} is generated using the Markovian embedding technique described in Section \ref{GLE_Simulation} for a particle with single-exponential memory $\Gamma(t) = \gamma e^{-t/\tau} / \tau$ in a harmonic potential, where $\tau$ is the memory time chosen as $\tau \omega_0=10$. In Figure~\ref{Figure_3}\textit{d}, we show spectra calculated using Eq.~\ref{linres_autocorr} for Markovian embedding simulations over a range of memory times for constant $\tau_m \omega_0=10$. The spectra exhibit a blue shift that is maximal for $\tau\omega_0 = 1$ and a continuous peak narrowing with increasing memory time $\tau$. These line shape changes, due to modifications of the velocity autocorrelation function, reflect non-Markovian effects \cite{Brunig_2022a}.

\subsection{Barrier-crossing reaction kinetics}\label{barrier_cross}

In Figures~\ref{Figure_3}\textit{e} and \ref{Figure_3}\textit{f}, we explore the consequences of memory effects on barrier-crossing reaction kinetics. For many systems, it is reasonable to assume the environment relaxes much faster than the barrier-crossing time scale. In these cases, one might be tempted to disregard memory effects entirely and use a theory assuming instantaneous friction, i.e., $\Gamma(t) = 2 \gamma \delta(t)$, such as the well-known Kramers' theory \cite{Kramers_1940}, which is suitable across a wide range of dynamic regimes. Here, we show that this time scale argument does not hold. In the overdamped limit, where mass can be neglected and there is a position-dependent friction profile $\gamma(x)$, the mean first-passage (MFP) time to reach $x_{\text{b}}$ starting at $x_{\text{a}}$ follows from the Smoluchowski equation for Markovian diffusion in a free energy profile $U(x)$ as
\begin{marginnote}[]
\entry{Kramers theory}{The first reaction rate theory to correctly account for the role of friction and inertia in predicting mean energy-barrier crossing times.}
\end{marginnote} 
\begin{equation}\label{tau_Exact}
\tau_{\rm{MFP}}^{\rm{\text{Mar}}} = \frac{1}{k_{\rm{B}}T}\int\limits_{x_\text{a}}^{x_\text{b}} dx \gamma(x)\text{e}^{ U(x)/k_{\rm{B}}T }
 \int\limits_{-\infty}^{x} dy\text{e}^{- U(y)/k_{\rm{B}}T}.
\end{equation}
This formula is particularly useful for systems with low energy barriers, since the Kramers theory is known to be inaccurate in this limit. The limitations of different reaction rate theories in the presence of non-Markovian friction can be assessed by analysis of the GLE, as demonstrated for the prominent example of pair reaction dynamics in a solvent \cite{Brunig_2022d}. In Figure~\ref{Figure_3}\textit{e}, we show a typical trajectory from a Markovian embedding simulation (Section \ref{GLE_Simulation}) with memory $\Gamma(t) = \gamma e^{-t/\tau}/\tau$. The external potential $U(x) = U_0\big[(x/L)^2-1\big]^2$ is a double-well with barrier height $U_0 = 3k_{\text{B}}T$. The simulations are performed in the overdamped limit ($\tau_{\text{m}}/\tau_{\text{D}}=0.001$) over a wide range of memory times, including the effective Markovian limit $\tau/\tau_{\text{D}} \rightarrow 0$. Reaction times are given as mean first-passage times $\tau_{\text{MFP}}$, evaluated from extended simulations \cite{Kappler_2018, Zhou_arXiv_2024}. Here, $\tau_{\text{D}} = \gamma L^2/k_{\text{B}}T$ represents the diffusion time scale, i.e., the time taken for the particle to diffuse over a characteristic distance $L$ in the absence of a barrier. In the Markovian limit, $\tau_{\text{MFP}}$ approaches the exact Markovian prediction Eq.~\ref{tau_Exact} for constant friction $\gamma(x) = \gamma$. As memory times are increased, the reaction times accelerate compared to the Markovian limit. This speed-up regime is predicted by the well-known Grote-Hynes theory \cite{Grote_1980} and was shown recently to be the dominant regime for many fast-folding proteins \cite{Dalton_2023, Dalton_PRL_2024}. However, for memory times beyond intermediate values $\tau > 0.1\tau_{\text{D}}$, reaction times increase rapidly, following quadratic scaling $\tau_{\text{MFP}}\sim \tau^2$. The Grote-Hynes theory, which assumes time-scale separation between the relaxation of the environment and the barrier-crossing process, is known to break down for systems with long enough relaxation times \cite{Straub_1986, Kappler_2018}, and cannot predict the memory-induced slow-down regime. Later theories, such as Pollak, Grabert, and H\"anggi theory, do capture this regime \cite{Pollak_1989}. It is important to note that throughout both the reaction speed-up and slow-down regimes, the memory times are much smaller than barrier crossing times, dispelling the requirement that $\tau$ should be equal to, or larger than, $\tau_{\text{MFP}}$ in order to influence barrier crossing dynamics, which follows from a naive time-scale separation argument.

In recent years, we have seen the development of an alternative approach for predicting barrier-crossing times in systems with memory. The interpolating crossover formulas \cite{Kappler_2018, Kappler_2019, Lavacchi_2020, Lavacchi_2022} interpolate between the low friction and intermediate-to-high friction regimes of Kramers theory for arbitrary mass, including an empirical crossover term between the two limiting behaviors. Non-Markovian effects are introduced through effective friction and mass terms, derived from the two-point correlation of a non-Markovian system in a harmonic well. For a memory kernel of the form $\Gamma(t) = \gamma e^{-t/\tau}/ \tau$, the equation becomes \cite{Kappler_2019}
\begin{equation}\label{tau_Heur}
\begin{split}
\frac{\tau_{\text{MFP}}}{\tau_{\text{D}}} =
\frac{ e^{U_0/k_{\text{B}}T}}{U_0/(k_{\text{B}}T)}
\left[  \left(  \frac{\tau_m}{\tau_{\text{D}}} + \frac{4U_0}{k_{\text{B}}T} \frac{\tau^2}{\tau^2_{\text{D}}} \right) 
+\frac{\pi }{2 \sqrt{2}}\frac{1}{1+  10 U_0\tau/( k_{\text{B}}T\tau_{\text{D}} )} + 2  \sqrt{ \frac{U_0\tau_m}{k_{\text{B}}T\tau_{\text{D}}}}\right],
\end{split}
\end{equation}
where the first term dominates under low friction, the second term under moderate to high friction, and the last term accounts for the turnover. This formula can extend to multi-exponential-component memory kernels \cite{Kappler_2019, Lavacchi_2020}. Figure~\ref{Figure_3} shows that Eq.~\ref{tau_Heur} accurately models $\tau_{\text{MFP}}$ for the simulation parameters across all memory times.

\section{Applications of GLE modeling and memory-kernel extraction}\label{applications}

In this section, we detail three recent applications of GLE modeling and memory kernel extraction, spanning the nano-, micro-, and macroscale.

\subsection{The nanoscale: molecular friction in protein folding}\label{app_protein}

Molecular friction plays a central role in the study of protein folding \cite{Plotkin_1998}. It has been known for some time that protein folding times $\tau_{\rm{fold}}$ do not scale linearly with solvent viscosity $\eta$, which would be predicted from Kramers theory and from Eq. \ref{tau_Exact}. This has led to the broad adoption of the notion of internal friction effects \cite{Beece_1980, Doster_1983, Ansari_1992, Jas_2001, Hagen_2010, Soranno_2012, Borgia_2012}, where it is assumed that dissipative intramolecular interactions, unaffected by solvent viscosity, contribute significantly to the folding times. In general, the viscosity dependence of protein-folding times scale according to $\tau_{\rm{fold}} = \alpha\eta^{\beta} + \varepsilon$, where $\beta=1$ and $\varepsilon=0$ in the absence of internal friction and either $\beta<1$ or $\varepsilon>0$ when internal friction effects are present. Internal friction effects have been studied extensively with all-atom simulations, and various molecular origins have been proposed \cite{Schulz_2012, Echeverria_2014, DeSancho_2014, Zheng_2015, Daldrop_2018}; however, discussions have remained phenomenological without an explicit theory relating $\tau_{\rm{fold}}$ and $\eta$. Fundamentally, deviations from $\tau_{\rm{fold}} \sim \eta$ indicate deviations from either the Stokes-Einstein relationship or, for sufficiently overdamped systems, Kramers theory for activated barrier crossing processes $\tau_{\rm{fold}} \sim\gamma$, thus identifying friction as the key factor in internal friction effects. Without methods to directly evaluate friction, it has not been possible to test each of these relationships separately. Recently, memory kernel extraction techniques were applied directly to the rotation of selected torsion angles from small isomerizing molecules, demonstrating that deviations from $\tau_{\rm{fold}} \sim\eta$ must be ascribed to simultaneous violations of both the Stokes-Einstein relation and the overdamped Kramers relation \cite{Dalton_2024}. Such a detailed analysis is still lacking for proteins.

\begin{figure}[t!]
\includegraphics[width=5in]{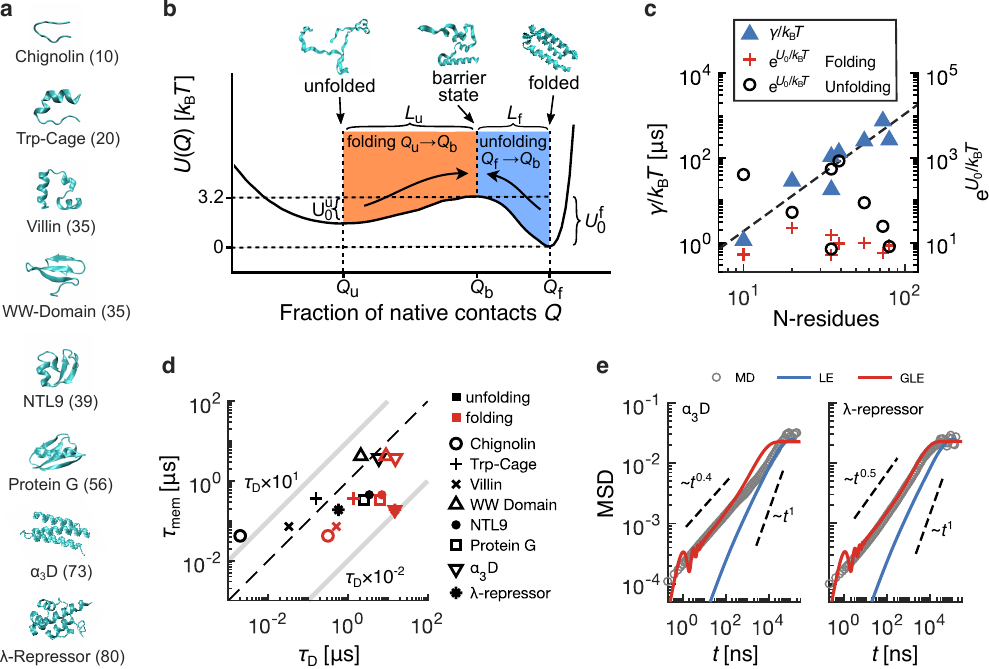}
\caption{GLE analysis of all-atom protein folding simulations. (\textit{a}) Native states of eight fast-folding proteins from extensive all-atom MD simulations \cite{Lindorff_2011}. (\textit{b}) Free energy profile for the fraction of native contacts reaction coordinate $Q$ for the $\alpha_3$D protein. Configuration snapshots show example unfolded ($Q_{\rm{u}}$), barrier-top ($Q_{\rm{b}}$), and folded ($Q_{\rm{f}}$) states. The folding and unfolding reactions are shown. (\textit{c}) The residue-number dependence of the friction coefficient $\gamma$ compared to the corresponding Arrhenius factor $\text{e}^{{U}_0/k_{\text{B}}T}$ for folding and unfolding transitions. Friction is divided by $k_{\rm{B}}T$ to account for unique system temperatures. A power law with an exponent $\sim2.8$ (black-dashed line) shows a positive correlation between friction and protein chain length. (\textit{d}) Memory-decay times $\tau_{\rm{mem}}$, calculated from the first moment of the memory kernel $\Gamma(t)$, compared to the diffusion times ($\tau_{\text{D}} = \gamma |Q_{\rm{b}} - Q_{\rm{u}}|^2/k_{\text{B}}T$ for folding and $\tau_{\text{D}} = \gamma |Q_{\rm{b}} - Q_{\rm{f}}|^2/k_{\text{B}}T$ for unfolding). Light gray lines indicate the bounding regime for $\tau_{\rm{mem}}$ between $\tau_{\rm{D}}{\times}10^{-2}$ and $\tau_{\rm{D}}{\times}10^{1}$, where memory-induced barrier-crossing acceleration is predicted. (\textit{e}) Mean squared displacement (MSD) for the $\alpha_3$D and the $\lambda$-repressor proteins. All-atom MD simulation results are compared to a purely Markovian Langevin model (LE) and a non-Markovian GLE model. All results originally published in Dalton et al. \cite{Dalton_2023}}.
\label{Figure_4}
\end{figure}

To evaluate the friction experienced by a folding protein, one must first reduce the full atomic details of the protein and its solvating environment to a suitably chosen reaction coordinate. In all-atom simulations, one can explore reaction coordinates and pursue those that optimize statistical measures, such as the transition-path probability, evaluated via transition-path ensemble methods \cite{Best_2005}. In the experimental context, one does not have the luxury of choosing optimal reaction coordinates, and a common choice is the inter-residue separation distance tracked indirectly by fluorescence resonance energy transfer (FRET) \cite{Schuler_2008, Chung_2012, Chung_2013} or directly by single-molecule force spectroscopy (SMFS) \cite{Neupane_2016, Neupane_2016B}. Regardless of technique, projecting onto a low-dimensional reaction coordinate is a coarse-graining procedure that incurs time-dependent, coordinate-specific friction. In protein folding, it is typically assumed that memory friction decays rapidly compared to relevant timescales, such as folding durations, leading to the conclusion that folding can be described by memoryless reaction-kinetic models. This enables the determination of friction via fitting procedures \cite{Hummer_2005, Gopich_2009}. As we demonstrated earlier (Section~\ref{barrier_cross}), this reasoning is not always justified. Still, Markovian techniques are commonly used to model protein folding \cite{Munzo_1999, Best_2006, Best_2010, Zheng_2015b, Chung_2015}. However, recent investigations have shown that memory-dependent friction influences the long-timescale reaction kinetics of fast-folding proteins \cite{Dalton_2023, Dalton_PRL_2024} and $\alpha$-helix-forming polypeptide chains \cite{Ayaz_2021}.

In Figure~\ref{Figure_4}, we discuss a recent application of friction extraction techniques to extensive all-atom protein folding simulations. In Figure~\ref{Figure_4}\textit{a}, we show native states for eight fast-folding proteins selected from the seminal work by Lindorff-Larsen et al. \cite{Lindorff_2011}. Recently, memory-kernel extraction techniques were applied to these data, revealing the friction memory kernels associated with protein folding \cite{Dalton_2023, Dalton_PRL_2024}. In Figure~\ref{Figure_4}\textit{b}, we show the potential of mean force for the fraction of native contacts reaction coordinate $x=Q$ projected from the full atomic trajectory of the $\alpha_3$D fast-folding protein. The total simulation time for $\alpha_3$D is $707\;\mu \text{s}$, during which the protein completes 12 folding and unfolding events. Comparison of free energy landscapes for all eight proteins shows that barrier heights for folding and unfolding are not correlated with protein chain length (Figure~\ref{Figure_4}\textit{c}). The total friction acting on the reaction coordinate $\gamma$, however, is correlated, suggesting that friction likely plays an increasingly important role in determining protein folding times for larger proteins. Using the same fast-folding protein data set, $Q$ was previously argued to be the optimal reaction coordinate for capturing transition states \cite{Best_2013}, often considered an indication of minimal non-Markovian effects \cite{Berezhkovskii_2018}. Despite this, the memory time scales in Figure~\ref{Figure_4}\textit{d} for the population of proteins in Figure~\ref{Figure_4}\textit{a}, estimated as the first moment of the memory kernel via $\tau_{\rm{mem}} = \int_{0}^{\infty} t \Gamma(t)dt/\int_{0}^{\infty} \Gamma(t)dt$, satisfy $1{\times}10^{-2} < \tau_{\text{mem}}/\tau_{\text{D}} < 1{\times}10^{1}$, which is the regime in which memory effects are known to accelerate barrier-crossing reaction times \cite{Kappler_2018}. Results show that the folding and unfolding times for these proteins are accelerated up to 10 times compared to Markovian predictions from Eq.\ref{tau_Exact}, and that the most accurate model for predicting protein folding and unfolding times is the multi-exponential extension of Eq.\ref{tau_Heur}, confirming that memory-dependent friction influences protein folding reaction kinetics.

Finally, when one analyzes the MSDs for this set of fast-folding proteins, one finds other hallmarks of memory friction effects. In Figure~\ref{Figure_4}\textit{e}, we show the MSDs from MD (grey circles) for the $\alpha_3$D and $\lambda$-repressor proteins. The plateau behavior at the longest timescale is due to the confinement of $Q$, which precludes long-time linear scaling, i.e., diffusive behavior. The intermediate regimes exhibit subdiffusive dynamics with exponents 0.4 and 0.5, characteristic of non-Markovian systems \cite{Goychuk_2009, Ayaz_2021}. Such subdiffusivity is often attributed to energy-landscape barriers. However, memoryless Langevin simulations performed on the extracted free energy profiles (blue lines) are diffusive for all time scales shorter than the confinement timescale. For multi-exponential Markovian embedding GLE simulations (see Section \ref{GLE_Simulation}), parameterized by the memory kernels extracted for the $\alpha_3$D and $\lambda$-repressor proteins, the MSDs (red lines) agree well with the simulation results.

\subsection{The microscale: classification of cell motion}\label{app_organism}

\begin{figure}[b!]
\includegraphics[width=5in]{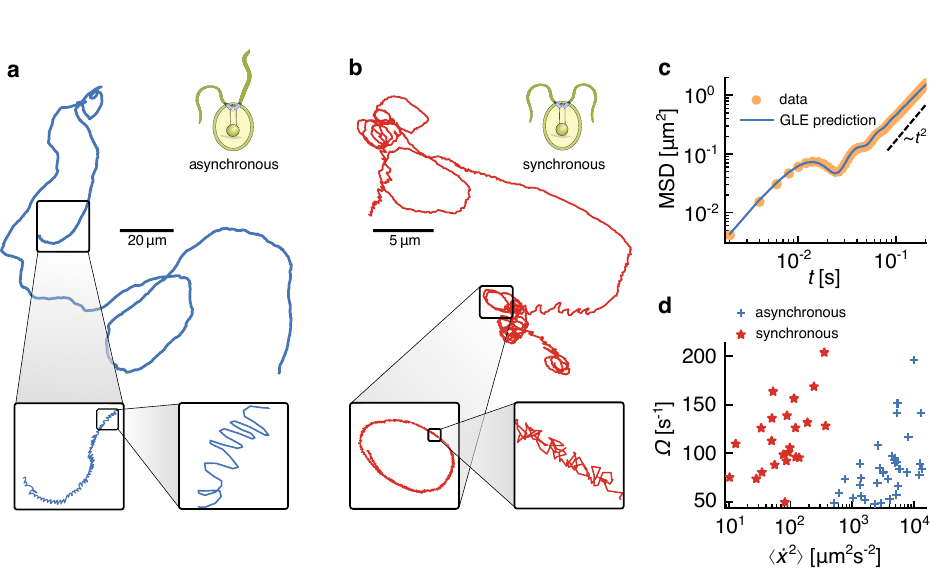}
\caption{GLE-based cell motion classification scheme. (\textit{a}) The swimming path of a \textit{Chlamydomonas reinhardtii} algal cell with asynchronous flagella beating. The duration of the trajectory is $\sim 50\;\rm{s}$, and the magnification focuses on short time scales of the motion. The oscillatory cell motion due to flagella beating can be seen at the highest resolution. (\textit{b}) A swimming path of a cell with synchronous flagella beating. The trajectory duration is also $\sim 50\;\rm{s}$. (\textit{c}) The mean squared displacement for a single algal cell with asynchronous flagella beating (orange points) shows oscillating motion on 0.01 s to 0.1 s time scales. The cell's directed motion shows a ballistic regime for time scales above 0.1~s. A theoretical prediction based on the GLE is also shown (blue line). (\textit{d}) Cell-type classification based on the extracted GLE parameters. The oscillation frequency $\Omega$, evaluated by fitting $\bar{\Gamma}(t) = 2 a\delta(t) + b e^{-t/\tau}\cos (\Omega t)$ (see Eq.~\ref{GLE_General}) to the memory kernels of single cells, is plotted against the velocity variance $\langle \dot{x}^2 \rangle$ for each cell. An unbiased cluster analysis reveals two distinct populations, corresponding to the swimming modes shown in (\textit{a}) and (\textit{b}). Results originally published in Klimek et al. \cite{klimek2024data}} 
\label{Figure_5}
\end{figure}

Living organisms are active, non-equilibrium systems. 

Therefore, for many processes in living systems, the equilibrium GLE models discussed so far may not suitably describe their dynamics. However, for studying long-time-scale motility and search dynamics, which are intrinsically stochastic processes, GLE models can indeed be applied \cite{Netz_2024,klimek2024data}. Various models have been proposed to describe the motion of different organisms \cite{bartumeus2005animal, johnson2008continuous, dieterich2008anomalous}. Examples include using Lévy walks to model wandering albatross flights \cite{edwards2007revisiting}, and the run-and-tumble or active Ornstein-Uhlenbeck models to describe bacterial cell motion \cite{tailleur2008statistical,martin2021statistical}. Some of these models are special cases of the GLE \cite{Mitterwallner_2020, Mitterwallner2020b}. Why certain organisms exhibit specific dynamic patterns are unclear. However, in some cases, there is likely an evolutionary pressure to search efficiently for resources \cite{viswanathan2011physics, klimek2022optimal}. Analyzing trajectories with the GLE reveals that different propulsion mechanisms lead to significantly different memory kernels. For example, human breast cancer cells show a negative memory component, leading to an extended ballistic regime in the MSD, while slime mold cells exhibit exponentially decaying memory \cite{li2011dicty}. Mouse fibroblasts and some other cells can be modeled by Markovian diffusion \cite{gail1970locomotion, wright2008differential}.

In addition to modeling organism motion, the GLE, in combination with memory kernel extraction techniques, can be used as the basis for a scheme to classify cell motion. In Figure~\ref{Figure_5}, we show swimming trajectories of two \textit{Chlamydomonas reinhardtii} algal cells confined between glass plates, recorded using video microscopy. Analysis of the flagellar motion of individual cells reveals two distinct swimming modes \cite{mondal2021strong}. The first mode, depicted in Figure~\ref{Figure_5}\textit{a}, is characterized by asynchronous flagella beating. This asynchronous beating leads to cell body wobbling, manifesting as short-time-scale oscillations observed in the finely magnified inset. The second distinct swimming mode is characterized by a synchronous breast-stroke motion of the flagella, as depicted in Figure~\ref{Figure_5}\textit{b}. On the shortest time scale, this synchronous beating leads to a back-and-forth motion. For both modes, the cells exhibit directed motion on intermediate time scales and diffusive behavior on long time scales. In Figure~\ref{Figure_5}\textit{c}, we show the MSD for a single asynchronously beating alga. The MSD resolves the oscillations from the flagella motion on 0.01 s to 0.1 s time scales, and above 0.1 s, the directed motion manifests as ballistic MSD ($\sim t^2$). The memory kernel, shown in Figure~\ref{Figure_2}\textit{c}, which has been extracted from the trajectory in Figure~\ref{Figure_5}\textit{a}, is well described by $\bar{\Gamma}(t) = 2 a\delta(t) + b e^{-t/\tau}\cos (\Omega t)$ \cite{klimek2024data}, which satisfies a general form that can be used to derive predictions for various two-point correlation functions, including the MSD and VACF. Such a prediction for the MSD, parameterized by fitting the memory kernel given in Figure~\ref{Figure_2}\textit{c}, agrees perfectly with the data in Figure~\ref{Figure_5}\textit{c}. Such a fitting scheme can be applied to parameterize the individual memory kernels evaluated for a population of cells. In Figure~\ref{Figure_5}\textit{d}, we show a scatter plot of the oscillation frequency $\Omega$ versus the mean-squared velocity $\langle \dot{x}^2 \rangle$, collected over a population of single cells. The oscillation frequency $\Omega$ is positively correlated with the mean-squared velocity, indicating that faster flagella beating leads to faster swimming. Applying an unbiased cluster analysis to the complete space of extracted GLE model parameters reveals two distinct clusters that coincide with the groups of synchronously beating and asynchronously beating cell types. Since the friction kernel is sensitive to the short time scale dynamics of algal motion, the GLE can be used for single-cell parameter extraction, which can then be used for cell-type classification. Due to the generality of this data-driven GLE approach, there is great potential for application to many other kinds of organism motion or for classifying cell properties, such as elasticity or size.

\subsection{The macroscale: predicting weather patterns}\label{app_weather}

\begin{figure}[b!]
\includegraphics[width=5in]{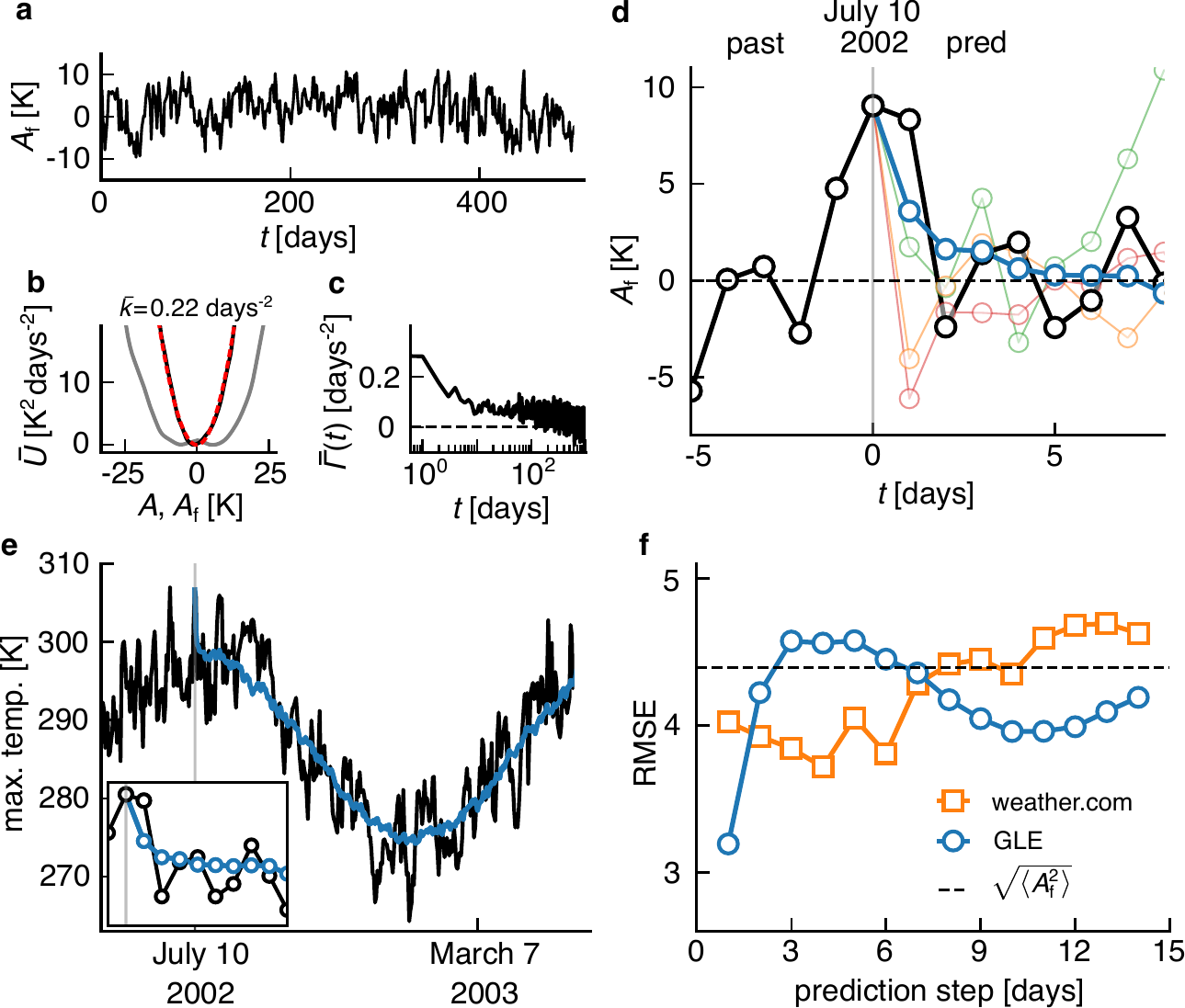}
\caption{Predicting weather patterns using the GLE \cite{Kiefer_2024}. (\textit{a}) The filtered temperature trajectory $A_\text{f}(t)$ obtained from the daily maximal temperature $A(t)$ in Figure~\ref{Figure_1}\textit{f} by removing the seasonal and slow components. (\textit{b}) Effective potential $\bar{U}$ (see Eq.~\ref{GLE_General}) of the unfiltered (gray) and filtered temperature data (black), compared with a harmonic potential $\bar{U} = \bar{k} A_\text{f}^2 / 2$ (dashed red). (\textit{c}) Memory kernel, $\bar{\Gamma}(t)$, extracted from $A_\text{f}(t)$. (\textit{d}) and (\textit{e}) Illustration of the GLE prediction scheme applied to the maximal temperature in Berlin-Tegel. (\textit{d}) The filtered temperature trajectory $A_\text{f}(t)$ over a 13-day period (black line). Thin colored lines show filtered trajectory predictions starting from July 10, 2002, using various random force realizations. The blue line shows the average over $N_\text{p}$ = 100 predictions. (\textit{e}) Prediction of the maximal temperature $A(t)$ (blue), generated by adding seasonal and slow components to the predicted $A_\text{f}(t)$, shown against actual data (black). The inset shows the first eight prediction steps. (\textit{f}) Comparison between the root-mean-squared error (RMSE) of a 14-day forecast, generated by the GLE model (blue), and predictions collected from the \textit{weather.com} website (orange). The black-dashed line denotes the standard deviation of $A_\text{f}(t)$, serving as a prediction benchmark.}
\label{Figure_6}
\end{figure}

The need to make sense of time-series data has implications across all spatial scales, from nanoscale to terrestrial scales. In meteorology and climatology, traditional models rely on basic physical principles such as conservation laws \cite{lorenc1986analysis}. In recent years, univariate and multivariate stochastic models \cite{franzke2015stochastic, watkins2021generalized} and machine-learning (ML) methods have been developed and applied to the study of weather patterns \cite{hochreiter1997long, zhang2003time, tseng2002combining, pathak2018model, raissi2018hidden, han2018solving}. Despite being universally applicable to such diverse data types, ML techniques also face considerable criticism. First, determining the exorbitant number of parameters necessary for accurate modeling incurs high computational costs for model training \cite{blum1997selection, al2015efficient}. Second, ML is often viewed as a black box, and the lack of interpretability of the learned rules is often unsatisfactory for applications in the natural sciences. The GLE is a general stochastic model and has been applied to financial and meteorological data \cite{schmitt_analyzing_2006, hassanibesheli2020reconstructing} and has been used to predict stochastic time-series data \cite{chorin_optimal_2000}. In time series analysis, the data are often decomposed into long-term trends, such as seasonal modes, and the remaining short-term components, which are often neglected or treated as stochastic \cite{zhang2003time, petropoulos2022forecasting}. This ad hoc decomposition makes the equation governing the dynamics of the remaining short-term components ill-defined. Here, we treat the raw signal $x(t) = A(t)$ using convolution filtering \cite{Kiefer_2024} according to 
\begin{equation} 
\label{eq:filtered_var}
A_\text{f}(t) = \int_{-\infty}^\infty dt' A(t-t') f(t').
\end{equation}
By constructing a filter function $f(t)$ that is a combination of low- and band-pass filters, the observable $A(t)$ can be exactly decomposed as $A(t) = A_\text{f}(t) + A_\text{lp}(t) + A_\text{bp}(t)$, where $A_\text{f}(t)$ is the fast-filtered component, $A_\text{lp}(t)$ is the slow, low-pass-filtered trend, and $A_\text{bp}(t)$ is the periodic, band-pass-filtered component. The slow and oscillating components $A_\text{lp}(t)$ and $A_\text{bp}(t)$ can be treated with deterministic models or fitted, the fast-filtered component $A_\text{f}(t)$ in fact, can be modeled by a GLE of the form given in Eq.~\ref{GLE_General} \cite{Netz_2024b}. The memory kernel $\bar{\Gamma}(t)$ can be extracted from the filtered time series $A_\text{f}(t)$ using previously discussed methods to predict the future trajectory $A(t)$.
 
The daily maximal temperature trajectory $A(t)$ for Berlin-Tegel in Figure~\ref{Figure_1}\textit{f} exhibits pronounced seasonal variation typical for a continental European climate, as seen from the filtered complement $A(t)-A_\text{f}(t)$, shown by the red line in Figure~\ref{Figure_1}\textit{f}. The filtered signal $A_\text{f}(t)$, shown in Figure~\ref{Figure_6}\textit{a}, exhibits stochastic behavior with temperature fluctuations in the range of $\pm$10 K. The effective potential for $A(t)$, given by $\bar{U}(A) = -\langle\dot{A}^2\rangle \log \rho( A)$, is shown in Figure~\ref{Figure_6}\textit{b} as a gray line. The two minima, reflecting the mean winter and summer temperatures, indicate that $A(t)$ is significantly non-Gaussian. The effective potential for the filtered trajectory $A_\text{f}(t)$ is harmonic (black line in Figure~\ref{Figure_6}\textit{b}), given by $\bar{U}(A_\text{f} )=\bar{k} A_\text{f}^2/2$, with $\bar{k} = \langle \dot{A}_\text{f}^2 \rangle/\langle A_\text{f}^2 \rangle = 0.22$ days$^{-2}$. This indicates that $A_\text{f}(t)$ is an effectively Gaussian stochastic variable. Therefore, the GLE in Eq.~\ref{GLE_General} is a suitable model for $A_\text{f}(t)$  \cite{Netz_2024,klimek2024data}. The memory kernel $\bar{\Gamma}(t)$, extracted from $A_\text{f}(t)$ (shown in Figure~\ref{Figure_6}\textit{c}), decays over many days, clearly revealing the non-Markovian nature of the daily temperature variation. To illustrate the GLE's predictive application, we arbitrarily set July 10, 2002, as the prediction date. To illustrate the process, we generate 100 realizations of the discrete future random force $\bar{F}_\text{R}$, which is conditioned on the history of $\bar{F}_\text{R}$, obtained by inversion of the GLE \cite{carof_two_2014}. In Figure~\ref{Figure_6}\textit{d}, three representative instances of the stochastic predictions of $A_\text{f}$ are shown. These predictions exhibit fluctuations similar to the actual filtered signal. The mean prediction, averaged over all instances, decays after a few days. After adding the filter complement $A(t)-A_\text{f}(t)$, we compare the full prediction of $A(t)$ from July 10, 2002 on with the actual temperature in Figure~\ref{Figure_6}\textit{e}. We quantify the accuracy of the prediction using the root-mean-squared error, $\text{RMSE}(j\Delta t) = \sqrt{\sum_{i=1}^{N_\text{s}} \bigl[ A(t_i + j\Delta t) - {A}_\text{pred}(t_i + j\Delta t)\big]^2/N_\text{s}}$, for $N_\text{s}=60$ predictions with randomly selected start times between March 9, 2020, and May 25, 2020. ${A}(t_i + j\Delta t)$ denotes the actual temperature trajectory, while ${A}_\text{pred}(t_i + j\Delta t)$ denotes the prediction starting at time $t_i$, with $i=1,2,\dots,N_\text{s}$. In Figure~\ref{Figure_6}\textit{f}, the GLE forecasts are more accurate than commercial \textit{weather.com} forecasts for single-day predictions and perform comparably overall. Additionally, GLE prediction methods are also much faster than state-of-the-art machine-learning techniques. This demonstrates the potential usefulness of predictions of complex data with the GLE.

% At this stage, $\Gamma(t)$ and $F_\text{R}(t)$ are analyzed for non-Markovian and non-Gaussian effects and checked whether the data contains predictable features. 

% Summary Points
%\begin{summary}[SUMMARY POINTS]
%\begin{enumerate}
%\item Summary point 1. Here, we can include some summary points about the review
%\item Summary point 2. Note that most people do not do this...
%\end{enumerate}
%\end{summary}

% Future Issues

\begin{summary}[SUMMARY POINTS]
\begin{enumerate}
\item We describe how the GLE can be used to analyze equilibrium and non-equilibrium time-series data from different disciplines. We discuss methods to derive GLEs, extract GLE parameters from time-series data, and simulate the GLE. 
\item For general model systems, we show that memory effects significantly modify diffusional behavior (described by MSD), vibrational behavior (described by the velocity-autocorrelation function), and barrier-crossing time in a bistable potential. 
\item The usefulness of the GLE approach is demonstrated with examples such as protein-folding dynamics, motility patterns of single cells, and weather data analysis and prediction.
\end{enumerate}
\end{summary}

% Future Issues
\begin{issues}[FUTURE ISSUES]
\begin{enumerate}
\item Robust methods for extracting GLE parameters from noisy and time-discretized experimental data need improvement \cite{Tepper_2024}.
\item The derivation of GLEs that include non-linear friction effects and can be simulated by efficient embedding techniques presents an open issue \cite{Ayaz_2022, AyazTurkish2022}.
\item The accurate prediction of non-Markovian reaction kinetics in arbitrary free energy landscapes is relevant for protein folding and chemical reaction kinetics but is not 
possible with current analytical reaction rate theories. 
\item The derivation and characterization of GLEs applicable to systems that are far from equilibrium even in their stationary state is an active research field \cite{Schilling_2022, Netz_2024, Schilling_2021}. Methods to detect and characterize non-equilibrium properties of systems based on time series data are particularly needed \cite{Netz_2023b}.
\end{enumerate}
\end{issues}

%Disclosure
\section*{DISCLOSURE STATEMENT}
The authors are not aware of any affiliations, memberships, funding, or financial holdings that might be perceived as affecting the objectivity of this review. 

% Acknowledgements
\section*{ACKNOWLEDGMENTS}

The work was supported by the European Research Council (ERC) Advanced Grant 835117 NoMaMemo, the Deutsche Forschungsgemeinschaft (DFG) Grant No. SFB 1449 "Dynamic Hydrogels at Biointerfaces", and the Deutsche Forschungsgemeinschaft (DFG) Grant No. SFB 1078 "Protonation Dynamics in Protein Function".

% \bibliography{/Users/dalton/Documents/Mendeley_Bibtex/Freie_Uni.bib}

% ----------------------------------------------------------------------------------------

% To generate a new .bib file from Freie_Uni source…

% In terminal, go to the folder. For Article_File.tex, run the following (assuming the manuscript is calling /Mendeley/Freie_Uni.bib):

% pdflatex Article_File.tex
% bibtex Article_File.aux
% pdflatex Article_File.tex
% pdflatex Article_File.tex

% bibexport -o bib_file.bib Article_File.aux

% Then, you can call output.bib in the Article_File.tex file

\bibliography{bib_file.bib}

% ----------------------------------------------------------------------------------------

\end{document}